\documentclass[final,5p,times,twocolumn]{elsarticle}

\usepackage{graphicx}
\usepackage{amssymb}
\usepackage{mathrsfs}
\usepackage{lineno}
\usepackage{listings}

\usepackage{amsmath,amsfonts,amsthm} % Math packages
\usepackage{xcolor}
\usepackage{enumitem}
\usepackage{booktabs}

\usepackage{threeparttable}
\usepackage{hyperref}

\definecolor{mygreen}{rgb}{0,0.6,0}
\definecolor{mygray}{rgb}{0.5,0.5,0.5}
\definecolor{mymauve}{rgb}{0.58,0,0.82}
\biboptions{square}

\bibpunct{(}{)}{;}{a}{}{,} % to follow the A&A style - 

\journal{astronomy \& computing}

\begin{document}

\begin{frontmatter}

%% Title, authors and addresses
\title{GriSPy: A Python package for Fixed-Radius Nearest Neighbors Search}
% About the design of image registration Python module: Astroalign

%% use the tnoteref command within \title for footnotes;
%% use the tnotetext command for the associated footnote;
%% use the fnref command within \author or \address for footnotes;
%% use the fntext command for the associated footnote;
%% use the corref command within \author for corresponding author footnotes;
%% use the cortext command for the associated footnote;
%% use the ead command for the email address,
%% and the form \ead[url] for the home page:
%%
%% \title{Title\tnoteref{label1}}
%% \tnotetext[label1]{}
%% \author{Name\corref{cor1}\fnref{label2}}
%% \ead{email address}
%% \ead[url]{home page}
%% \fntext[label2]{}
%% \cortext[cor1]{}
%% \address{Address\fnref{label3}}
%% \fntext[label3]{}

%% use optional labels to link authors explicitly to addresses:
%% \author[label1,label2]{<author name>}
%% \address[label1]{<address>}
%% \address[label2]{<address>}

\author[iate,oac,famaf]{Martin Chalela}%\thanks{E-mail: mchalela@unc.edu.ar}}
\author[iate,oac,famaf]{Emanuel Sillero}
\author[iate,oac,famaf]{Luis Pereyra}
\author[utn]{Mario Alejandro Garcia}
\author[cifasis,iate]{Juan B. Cabral}
\author[iate]{Marcelo Lares}
\author[iate,oac]{Manuel Merch\'{a}n}

\address[iate]{
   Instituto de Astronom\'ia Te\'orica y Experimental -
   Observatorio Astron\'omico de C\'ordoba (IATE, UNC--CONICET),
   C\'ordoba, Argentina.}
\address[oac]{
    Observatorio Astron\'{o}mico de C\'{o}rdoba, Universidad Nacional de C\'{o}rdoba, Laprida 854, X5000BGR, C\'{o}rdoba, Argentina
}
\address[famaf]{
	Facultad de Matem\'atica, Astronom\'{\i}a y F\'{\i}sica,
    Universidad Nacional de C\'ordoba (FaMAF--UNC)
	Bvd. Medina Allende s/n, Ciudad Universitaria,
    X5000HUA, C\'ordoba, Argentina 
}
\address[utn]{
    Universidad Tecnol\'ogica Nacional, Facultad Regional C\'ordoba (UTN--FRC), Maestro M. Lopez esq. Cruz Roja Argentina, Ciudad Universitaria - C\'ordoba Capital
}
\address[cifasis]{
   Centro Internacional Franco Argentino de Ciencias de la
   Informaci\'on y de Sistemas (CIFASIS, CONICET--UNR),
   Ocampo y Esmeralda, S2000EZP,
   Rosario, Argentina.}

\begin{abstract}
We present a new regular grid search algorithm for quick fixed-radius nearest-neighbor lookup developed in Python. This module indexes a set of k-dimensional points in a regular grid, with optional periodic conditions, providing a fast approach for nearest neighbors queries. In this first installment, we provide three types of queries: \textit{bubble}, \textit{shell} and the \textit{nth-nearest}. For these queries we include three different metrics of interest in astronomy, namely, the \textit{euclidean}, the \textit{haversine} and the \textit{Vincenty}, the last two in spherical coordinates. We also provide the possibility of using a custom distance function. This package results particularly useful for large datasets where a brute-force search turns impractical.

\end{abstract}

\begin{keyword}
Data mining: Nearest-neighbor search; Methods: Data analysis; Astroinformatics; Python Package

%% MSC codes here, in the form: \MSC code \sep code
%% or \MSC[2008] code \sep code (2000 is the default)

\end{keyword}
\end{frontmatter}

%%
%% Start line numbering here if you want
%%
%\linenumbers

% =============================================================================
% Section Intro
% =============================================================================

\section{Introduction}

The nearest neighbor search (NNS) problem can be defined as follows: given a set $P$ of $n$ points defined in the multidimensional space $X$ with distance function $D$, run an algorithm that, given a query point $q \in X$, finds the point $\min_{p \in P} D(q,p)$. This problem arises in a wide range of scientific fields, including machine learning, robotics , chemistry, astronomy and many other areas of application \citep[e.g.][]{Shakhnarovich2006,Teofili2019,Devlin2015,Calle-Vallejo2015}.

In the particular field of astronomy, the everyday increasing amount of observational and simulated data requires algorithms that can handle the computational demands. Most modern cosmological simulations consist of over $10^{10}$ particles, e.g. the Illustris Project \citep{Springel2018, Vogelsberger2014}, the MultiDark Simulation \citep{Klypin2016} or the Millennium Simulation \citep{BoylanKolchin2009}, with the additional feature of being in a 3D box with periodic boundary conditions. Even smaller scale simulations may consist of $10^6$ particles. On the other hand, the observational community is also facing this problem thanks to large-scale sky surveys such as the Sloan Digital Sky Survey \citep{Alam2015} and the Dark Energy Survey \citep{Zuntz2018}, and will face even greater challenges with upcoming projects like the Large Synoptic Survey Telescope \citep{Ivezic2019}.

Several methods have been proposed for solving the NNS problem and according to their solution they can be broadly divided in \textit{approximate} or \textit{exact}. Approximate solutions are usually of interest when working with high dimensional datasets and they retrieve points that may fall outside the query radius by a given uncertainty parameter $\epsilon$, such that $D(q,p) \leq (1+\epsilon)D(q,p)^{*}$, where $D(q,p)^{*}$ is the true distance \citep{Maneewongvatana1999}.

The simplest and more direct solution to the problem is the \textit{brute force} method, which requires to compute the distance $D(q,p)$ for every point $p \in P$. The data structure required by this method is quite simple, mainly an array with the original set of points, and thus the memory and CPU overhead are very small. However, given that it performs every possible distance calculation, for large number of points this becomes computationally expensive and a different approach is needed. The most popular method is to apply a partitioning-indexing scheme to track the approximate location of points in the multidimensional space. Among the algorithms that apply this concept are the \textit{binary-tree} and \textit{cell techniques}. Binary tree methods iteratively divide the space into two nodes, or branches, in each iteration until a certain number of particles is reached. The overhead and construction time of the tree structures can be quite large but in exchange they offer fairly short query times. For a detailed review of binary-trees the reader is refered to the seminal works by \citet{Friedman1977} and \citet{Bentley1975}. On the other hand, cell techniques create a regular grid, or hypercube, in the multidimensional domain and through a simple math operation every point is assigned an integer lattice that points to its corresponding cell. A hash table can then be used for future queries, where the same math operation is applied to the query point to know which cell it belongs to. The distance to every point in the cell, and probably in the contiguous cells as well, has to be computed to return only the points that meet the query condition.

GriSPy adopts the cell technique approach to solve the NNS problem in a multidimensional space. This method provides the perfect particularities that can be exploited by the extremely efficient NumPy \citep{VanDerWalt2011} routines to handle large-sized arrays. Among the key features GriSPy provides, are the possibility of working with periodic boundary conditions, individual search radius for each query point in fixed-radius searches and minimum and maximum search radius for shell queries. This package can be a very useful tool in many areas that need an optimized solution for the NNS problem. In the field of astronomy, for example, GriSPy can be used in gravitational N-body simulations with periodic conditions, galaxy catalogues from large observational surveys, studies of the $\Sigma_5$ density parameter and its correlation with the evolution of galaxies, in tools to describe the statistical properties of the large scale structure of the Universe among many others.

% =============================================================================
% Section Description
% =============================================================================

\section{Description of the Algorithm}

\begin{figure}
\includegraphics[width=\columnwidth]{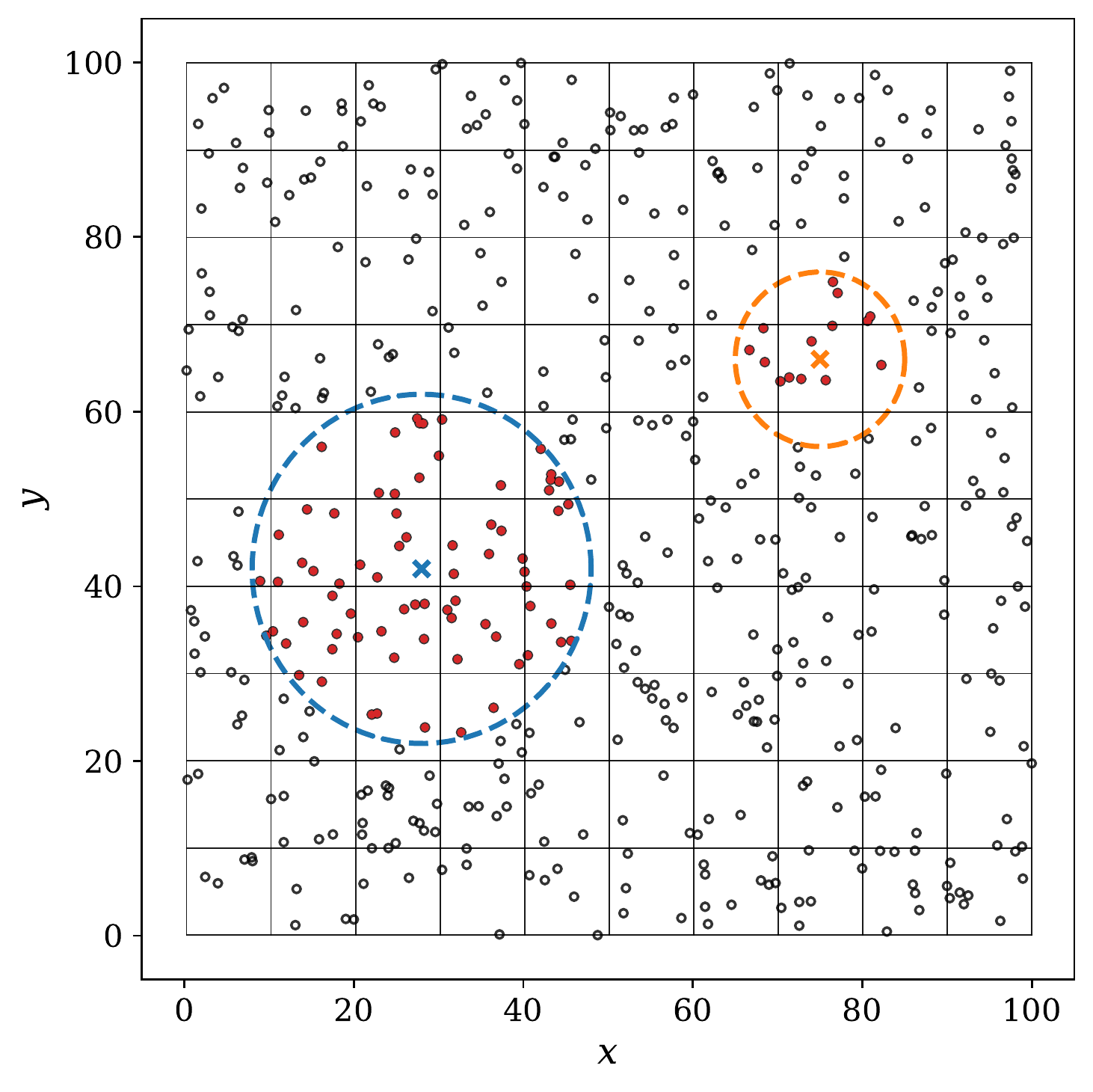}
\caption{Example of a 2D uniform distribution of points with a grid of $10 \times 10$ cells. Two centres with different search radius are plotted and their corresponding neighbors are marked as filled circles.}
\label{fig:grid}
\end{figure}

In this section we describe the details of the partitioning-indexing method for the construction of the axis-aligned grid and how we query for neighbors.

\subsection{Indexing}

Given an initial set of $k$-dimensional points, a regular grid of $N^k$ cells is built in the domain of the data. The minimum and maximum values of the grid in each dimension, i.e. the box walls, are those of the data itself, expanded with a small margin to avoid numerical leaks.

After the grid boundaries are defined, the coordinates of a data point, $x_i$, can be converted to grid coordinates using:

\begin{equation}
    g_i = \textrm{int}\left( N \, \frac{x_i - w_{l, \, i}}{w_{r, \, i} - w_{l, \, i}} \right), \quad  \, i = 1, 2, ..., k
\label{eq:grid_coord}
\end{equation}
where $w_{l, \, i}$ and $w_{r, \, i}$ are the left and right wall coordinates, respectively, for the coordinate $i$. Once the grid coordinates are computed, a hash table is created where the key is each cell coordinate and the value is a list containing the indices of every point within that particular cell. For GriSPy we implemented a Python dictionary using a tuple as key and a list as value.

\subsection{Searching}

Once the hash table is created, the query for neighbors within a given radius is straightforward. This is the basis of the ``bubble search'' method.
First, we extract the box of cells that contains the hyper-sphere using equation \ref{eq:grid_coord} and then keep only those cells touched by the hyper-sphere. We then retrieve every data point contained within those cells using the hash table and compute the distance to remove those points located outside the hyper-sphere:

\begin{equation}
    D(x_0, x) \leq r_{max}
\label{eq:bubble_dist}
\end{equation}

In the case of a shell query, i.e. points with a distance between a minimum ($r_{min}$) and maximum ($r_{max}$) radius, we remove from the distance computation those inner cells untouched by the minimum radius to exclude points that we know \textit{apriori} are outside the distance bounds. We then retrieve those points that meet the condition:

\begin{equation}
    r_{min} < D(x_0, x) \leq r_{max}
\label{eq:shell_dist}
\end{equation}

As a feature of GriSPy, a different radius can be provided for each centre in both types of queries.
The last type of query implemented is the $n$-th nearest neighbors. Given that it is not possible to know beforehand exactly how many neighboring cells need to be opened, we make an initial estimation using the length of a cell diagonal as the radius of a bubble query. Then this radius is used in iterative shell queries until the $n$-th nearest neighbors are found.

\subsection{Distance metrics}

To compute the distance between two points we implemented three metrics, for the first version of the package, that are of interest in astronomy. The \textit{euclidean} distance defined as:
\begin{equation}
    D(x_0, x) = \sqrt{ \sum_{i=1}^{k} (x_{0, i} - x_i)^{2} } \, \textrm{;}
\label{eq:euclid}
\end{equation}
and two distance functions defined on the surface of a unit sphere. In these cases the set of points and centres coordinates are two-dimensional and correspond to longitude and latitude, i.e. ($\lambda, \varphi$). One of them is the \textit{haversine} formula which determines the great-circle distance:
\begin{equation}
    D(x_0, x) = 2 \arcsin{ \sqrt{\, \sin^2 \left( \frac{\Delta\varphi}{2} \right) +
    \cos\varphi_0 \, \cos\varphi \, \sin^2 \left( \frac{\Delta\lambda}{2} \right)} }
\label{eq:haversine}
\end{equation}

The last distance function is the \textit{Vincenty} formula \citep{Vincenty1975} which solves numerical problems for very close points and antipodal points at the expense of more computing time. The general formula gives the distance between two points on the surface of an ellipsoid. However, we are interested in the case where the major and minor axes are equal. The distance function is then:
\begin{align}
    D(x_0, x) & = \arctan\frac{\sqrt{ E^2 + F^2 }}{G} \, \textrm{,} \\
    E & = \cos\varphi \, \sin(\Delta\varphi) \nonumber \\
    F & = \cos\varphi_0 \, \sin\varphi - \sin\varphi_0 \, \cos\varphi \, (\Delta\lambda) \nonumber \\
    G & = \sin\varphi_0 \, \sin\varphi + \cos\varphi_0 \, \cos\varphi \, (\Delta\lambda) \nonumber
\label{eq:vincenty}
\end{align}

\subsection{Periodicity}

\begin{figure}
\includegraphics[width=\columnwidth]{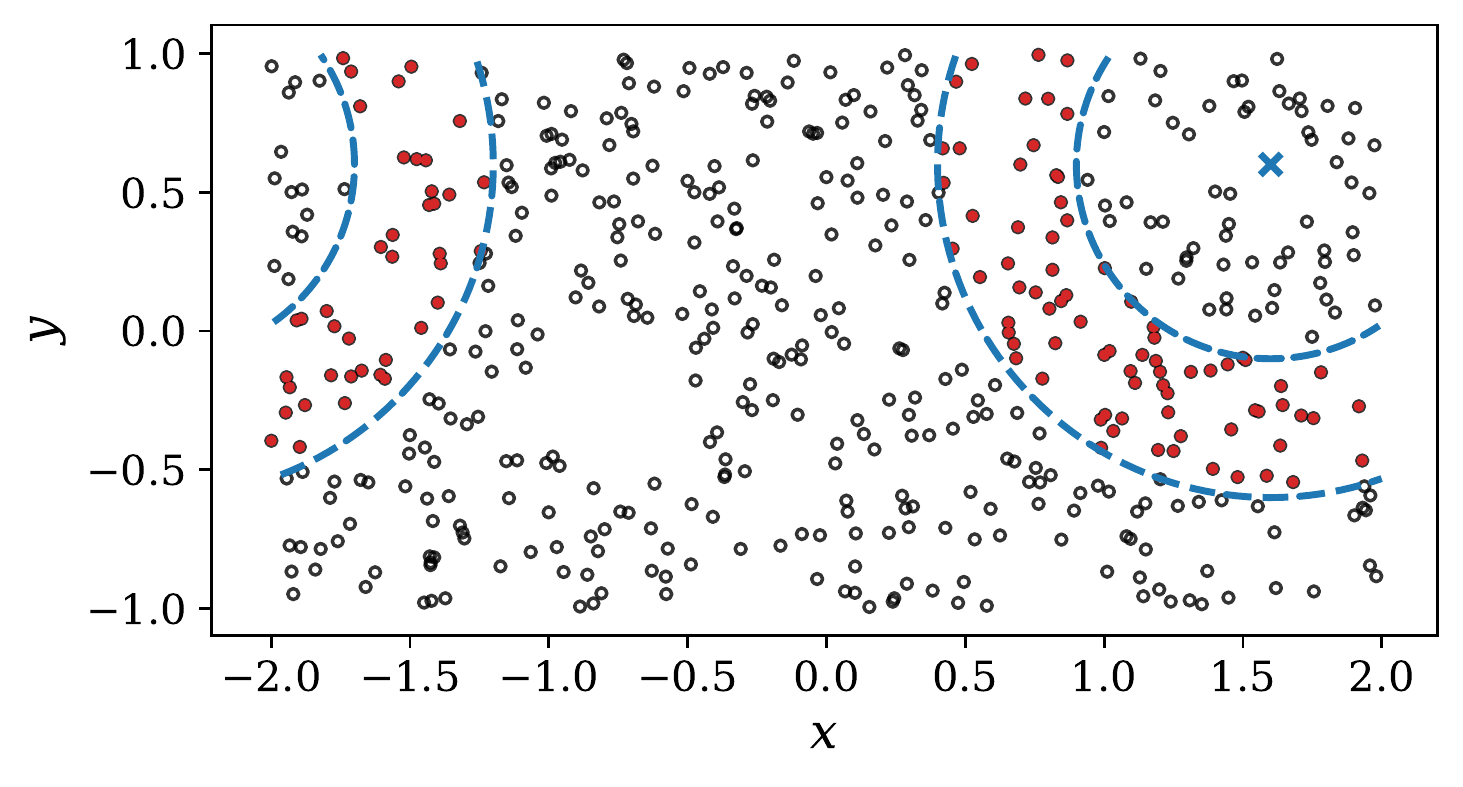}
\caption{Example of a shell query where the domain presents periodicity in one dimension. The corresponding neighbors are marked as filled circles.}
\label{fig:periodic}
\end{figure}

Periodicity is a key ingredient in many simulations, where it is beyond practical capabilities to simulate an extremely large box. Instead, a smaller, representative box, with periodic boundary conditions is used. Particles near the box walls experience the effects caused by the presence of a \textit{ghost} box that starts exactly where the main box ends.
When searching for neighbors of a centre with a search radius that extends beyond the box edge, the algorithm needs to retrieve points located on the opposite side of the box. To implement this behavior we create \textit{ghost} centres located at a distance $L_{box}$ (i.e. one box width) in the opposite direction as shown in Figure \ref{fig:periodic}. In GriSPy we implemented axis-independent periodic conditions, i.e. each dimension may or may not present periodic boundaries.

% =============================================================================
% Section Code details
% =============================================================================

\section{Technical details about the GriSPy package}
\label{sec:tech_details}
Throughout the entire implementation of GriSPy we make heavy use of NumPy \citep{VanDerWalt2011} vectorized methods and array broadcasting properties to achieve high performance. NumPy provides efficient implementation of numerical computations in a high-level language like Python but completely compiled in C, resulting in a significant speed improvement and in code that is both transparent and easy to maintain.

\subsection{User functionalities}
GriSPy is an object oriented package that exposes the main grid constructions as a \texttt{GriSPy()} class. In the configuration step the user provides the set of \textit{k}-dimensional points to be indexed, and optionally some other configuration parameters such as the periodicity conditions, the number of cells and the distance metric. Besides the three distance metrics provided by GriSPy, the user has the possibility of providing a callable custom distance function in the \texttt{metric} argument.

The instance of the GriSPy class has the following queries implemented as methods:

\begin{itemize}
    \item \texttt{bubble\_neighbors()}: find neighbors within a given radius. A different radius for each centre can be provided. Neighbors can be sorted by distance.
    \item \texttt{shell\_neighbors()}: find neighbors within given lower and upper radius. Different lower and upper radii can be provided for each centre. Neighbors can be sorted by distance.
    \item \texttt{nearest\_neighbors()}: find the \textit{n}-th nearest neighbors for each centre. Neighbors can be sorted by distance.
\end{itemize}

Also, the following method is available:

\begin{itemize}
    \item \texttt{set\_periodicity()}: optional periodic boundary conditions can be provided for each axis individually.
\end{itemize}

An in depth description of the methods parameters can be found in the documentation (see Section \ref{subsec:quality}).

\subsection{Application example} \label{subsec:example}

As a simple usage application, we show how to compute the two-point correlation function ($\xi$(r)) in a gravitational N-body simulation.
The spatial particle-particle autocorrelation function $\xi(r)$, measures the excess probability $dP$ with respect to a random distribution, that a particle will reside at a distance $r$ away from a another particle, in a volume element $dV$. This can be expressed as
\begin{equation*}
dP=\bar{n}[1+\xi(r)]dV,
\end{equation*}
where $\bar{n}$ is the mean number density of the simulation. A standard method to measure $\xi(r)$ is the Davis \& Peebles estimator \citep{DavisPeebles1983}, that consists on counting for each particle (centre), the number of neighbouring objects (tracers) found at different distance bins. The total number of neighbours per interval, $DD$, is then normalized by the number of pairs expected in a homogeneous distribution, $DR$. Finally, for each distance bin, the excess with respect to the unit of the stacked count is our estimator of the correlation function $\xi(r) = DD(r)/DR(r)-1$.

%with the Davis \& Peebles estimator \citep{DavisPeebles1983} in a gravitational N-body simulation
We use the last snapshot (redshift zero) of a dark matter only simulation of $512^{3}$ particles in a periodic box of $500 \, h^{-1} \mathrm{Mpc}$ side with cosmological parameters $\Omega_{m} = 0.258$, $\Omega_{\Lambda} = 0.742$, $h = 0.719$ and with a normalization parameter $\sigma_{8} = 0.796$. The simulation was evolved using the public version of GADGET-2 code \citep{Springel2005} and used in other works \citep[e.g.][]{Paz2011}. 

\begin{figure}
\includegraphics[width=\columnwidth]{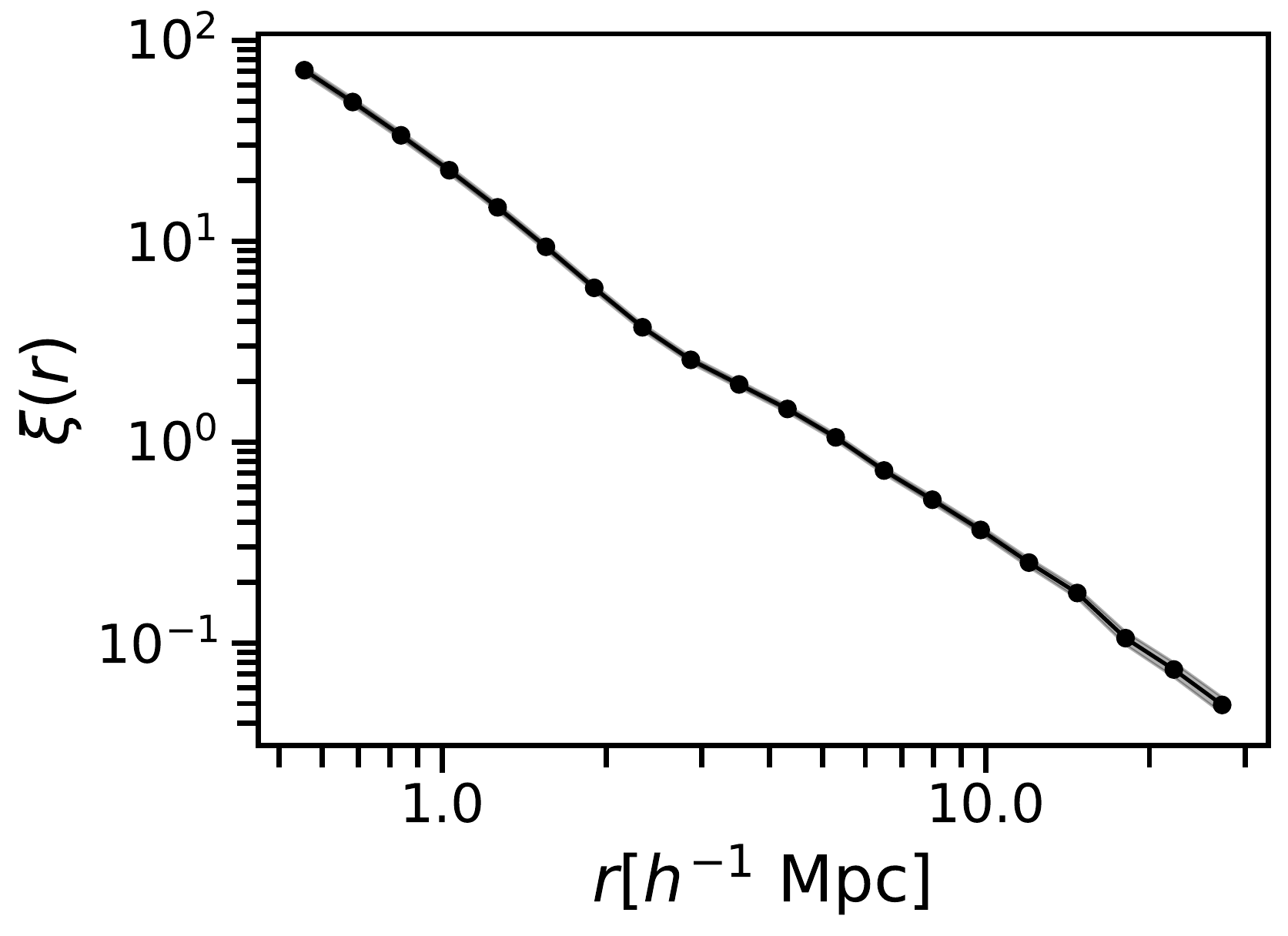}
\caption{Example of a two-point correlation function.
The black line with filled regions show the correlation function together with its errors based on a jackknife estimate.
} \label{fig:tpcf}
\end{figure}

In Figure \ref{fig:tpcf}, we present the resulting correlation function $\xi(r)$. The error estimations are obtained through a jackknife method.

\begin{lstlisting}[language=Python]
>>> import numpy as np

# import the class from the grispy package
>>> from grispy import GriSPy

# number of bins
>>> Nbins = 20  
>>> r_min, r_max = 0.5, 30.0 
>>> bins = np.geomspace(r_min, r_max, Nbins+1) 

# Box of width lbox, with periodic conditions
>>> lbox = 500.0
>>> periodic = {0: (0, lbox), 
...             1: (0, lbox),
...             2: (0, lbox)}

# Build GriSPy object
# Pos is the position array of shape=(N, 3)
#    where N is the number of particles
#    and 3 is the dimension
>>> gsp = GriSPy(Pos, periodic=periodic)

# Query Distances
>>> shell_dist, shell_ind =  gsp.shell_neighbors(
...     Pos, distance_lower_bound=r_min,
...     distance_upper_bound=r_max)

# Count particle pairs per bin
>>> counts_DD = np.zeros(Nbins)
>>> for ss in shell_dist:
...     cc, _ = np.histogram(ss, bins)
...     counts_DD += cc

# Compute the two-point correlation function
# with theoretical randoms
>>> npart = len(Pos)
>>> rho = npart / lbox**3
>>> vol_shell = np.diff(
...     4.0 * np.pi / 3.0 * bins**3)
>>> count_DR  = npart * rho * vol_shell

>>> xi_r = count_DD/count_DR - 1
\end{lstlisting}

Another interesting example is to provide GriSPy with a custom distance-metric. 

In our project parameters the metric is an arbitrary Python function that must take three arguments: \texttt{centre}, the position of a single centre; \texttt{targets}, the position of the points to which we want to calculate the distance from \texttt{centre}; and \texttt{dim}, the dimension of \texttt{centre} and \texttt{targets}. 
The return value of the function must be a NumPy array of the same length as \texttt{targets}, where the \textit{j-nth} element corresponds to the distance between \texttt{centre} and \texttt{targets$_j$}.

For example, if we wanted for some reason to implement a Hamming distance metric \citep{bookstein2002generalized}, taking advantage of the functionalities of the \textit{SciPy} \footnote{\url{https://docs.scipy.org/doc/scipy/reference/generated/scipy.spatial.distance.cdist.html}} distance package, we can write:

\begin{lstlisting}[language=Python]
>>> from scipy.spatial.distance import cdist
>>> def hamming(centre, targets, dim):
...    centre = centre.reshape((-1, dim))
...    d = cdist(
...        centre, targets, metric="hamming"
...    ).reshape((-1,))
...    return d

\end{lstlisting}

Then we can create the grid with the custom distance, and run the code as follows:

\begin{lstlisting}[language=Python]

>>> gsp = GriSPy(data, metric=hamming)
>>> ham_dist, ham_ind = gsp.bubble_neighbors(
...    centres, distance_upper_bound=10.)

\end{lstlisting}

\subsection{Quality assurance} 
\label{subsec:quality}
To ensure the proper software quality of the GriSPy package and the development process, we provide standard qualitative and quantitative metrics, in particular \textit{unit-testing} and \textit{code-coverage}, and endorse the PEP 8 style guide throughout the entire project. 

The purpose of unit-testing is to validate that the individual components of the software work as expected \citep{Jazayeri2007}. GriSPy is tested for Python versions 3.6, 3.7 and 3.8.  Code-coverage measures how much of the code is covered by the unit test suite, expressed as a percentage of executed sentences \citep{MillerMaloney1963}. Providing an exhaustive code-coverage prevents major parts of the code from being untested and ensures that fundamental errors have been properly handled. In the GriSPy project we provide four suites of unit-tests that evaluate different sections of the code, reaching 99\% of code-coverage. We use the \textit{pytest}\footnote{\url{https://pytest.org}} package in the test suite and  \textit{coverage.py}\footnote{\url{https://coverage.readthedocs.io}} to measure the code coverage. As we are interested in the maintainability of the project, we adopted the \textit{PEP 8 -- Style Guide for Python Code} \citep{PEP8} to improve the readability and consistency of the code by using the \textit{flake8}\footnote{\url{http://flake8.pycqa.org}} tool, which ensures that there are no deviations in style and will help minimize the "code-entropy" of future versions.

The complete source code is under the MIT-license \citep{MITLicense}, and available in a public repository\footnote{\url{https://github.com/mchalela/GriSPy}}. Changes and new versions committed to this repository are automatically tested with a continuous-integration service\footnote{\url{https://travis-ci.org/mchalela/GriSPy}}. Documentation is automatically generated from GriSPy docstrings and made public in the read-the-docs service\footnote{\url{https://grispy.readthedocs.io/en/latest/index.html}}.

At last, GriSPy is available for installation on the \textit{Python-Package-Index} (PyPI)\footnote{\url{https://pypi.org/project/grispy/}}. The interested user can install it via the command \texttt{pip install grispy}; and finally the project is registered in the ``Astrophysics Source Code Library'' (ASCL)\footnote{\url{https://ascl.net/code/v/2439}}\citep{Allen2015}, \citet{grispyASCL}.

% =============================================================================
% Benchmarking
% =============================================================================

\subsection{Benchmarking}  \label{subsec:bench}
As previously seen, the GriSPy algorithm can be divided in two steps: build and query. The time taken by each one of them to return results will highly depend on their respective input parameters. In order to asses their time performance we created a series of scenarios where key parameters are varied and the \textit{user time} is measured. 

Every input parameter has an impact on the time taken to return a given neighbors query. For example, if the parameter \texttt{sorted=True} is passed as an argument to \texttt{bubble\_neighbors()}, it will naturally take longer to return results because the neighbors will be ordered according to their distances. Given that build and query times (hereafter BT and QT, respectively) depend on a complicated way on every input parameter, we focus on those of most interest: the number of dimensions ($k$), the number of grid cells ($N_{cells}$), the number of data points ($N$) and the number of query centres ($N_{centres}$). Of these parameters the number of grid cells is the only one that can be modified to optimize the queries, the rest depend on the particular problem and most of the time can not be changed. For this reason we analyze the time dependence with respect to the number of cells, varying a given parameter and fixing the rest. This will also give us a helpful insight about the optimal choice of the default value. Two cases are considered for the analysis: a uniform random distribution and the N-body simulation used in the previous example (see Section \ref{subsec:example}).

\begin{figure*}
\includegraphics[width=\textwidth]{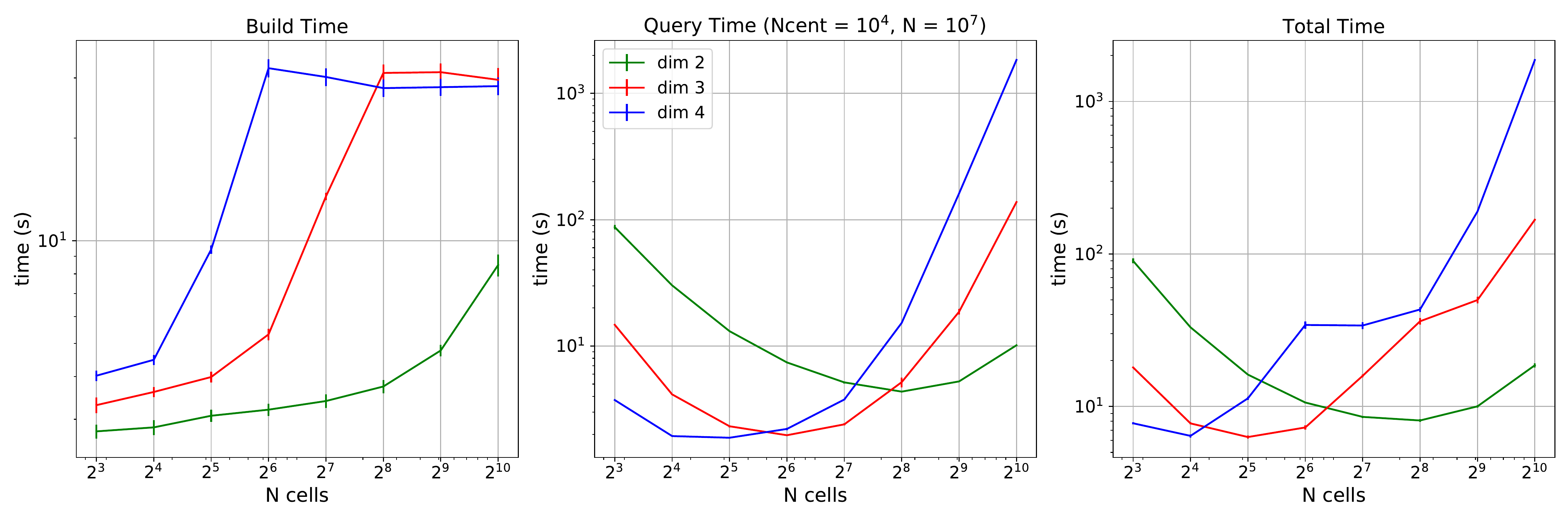}
\includegraphics[width=\textwidth]{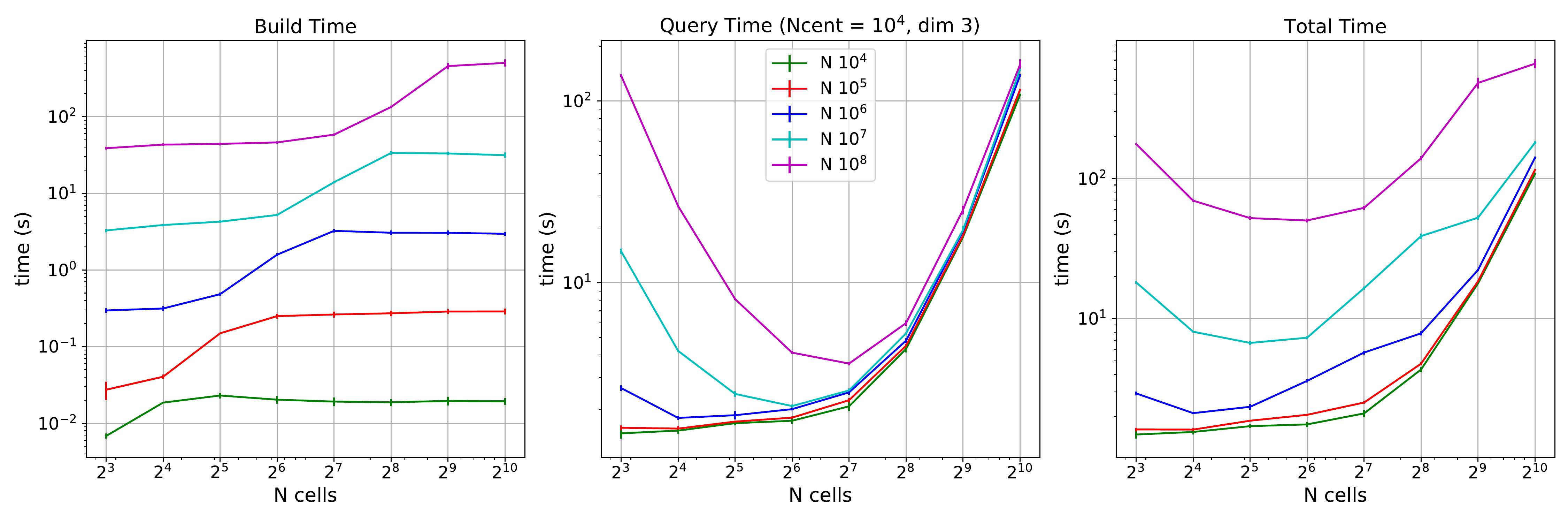}
\includegraphics[width=\textwidth]{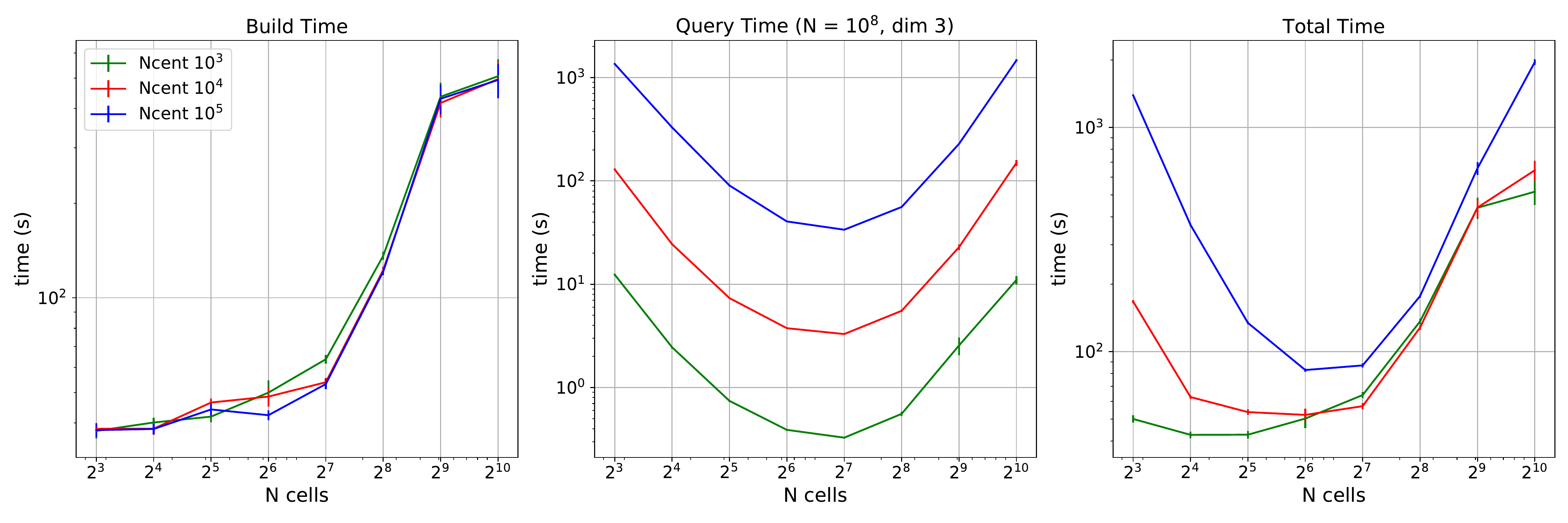}
\caption{Time benchmark of GriSPy in a uniform distribution of points using the \texttt{bubble\_neighbors} method with a search radii of 1\% of the box size. Each column represents the time spent in each step, from left to right: Build, Query and Total time. First row: varying dimension for fixed number of data-points ($N$) and centres ($N_{centres}$). Second row: varying number of data-points ($N$) for fixed dimension and number of centres ($N_{centres}$). Third row: varying number of centres ($N_{centres}$) for fixed dimension and number of data-points ($N$). In all cases the error bars are the standard deviation of 10 realizations.}
\label{fig:bench}
\end{figure*}

We first created a random uniform distribution with values in the range (0, 1) in each dimension. For the queries we used the \texttt{bubble\_neighbors()} method with a search radius of 0.01. All distances are computed with the \textit{euclidean} metric. In Figure \ref{fig:bench} we show the results of the analysis where the relation \textit{time vs. $N_{cells}$} is studied for three cases:

\begin{itemize}
    \item[(a)] \textbf{varying dimension ($k$) for fixed number of data-points ($N$) and centres ($N_{centres}$):} The BT increases for increasing number of cells, despite the dimension $k$. However, when the number of total cells ($N_{cells}^k$) approximates to $N$, i.e. one point per cell, the BT stops increasing. This is because GriSPy only indexes occupied cells, so increasing the number of cells beyond $N$ does not increase the BT. The QT shows a minimum value indicating the optimal $N_{cells}$ value for a given set of points and centres. This behaviour is the same for different $k$, but the minimum value shifts towards smaller $N_{cells}$ for higher dimensions. The total time (TT) shows the sum of both curves and the optimal value is clearer.

    \item[(b)] \textbf{varying number of data-points ($N$) for fixed dimension ($k$) and number of centres ($N_{centres}$):} The BT again stops increasing when the number of total cells approximates to $N$. We can also see that the curves are approximately separated by an order of magnitude, showing that the BT scales linearly with $N$. In the QT we see the minimum time shifts to larger $N_{cells}$ for increasing $N$. The QT then increases independently of $N$. This is the region where the number of points is $\sim1$ per cell and most of the time is consumed in the distance computation to grid centres. The TT shows the sum of both curves and how the optimal value shifts towards less $N_{cells}$.

    \item[(c)] \textbf{varying number of centres ($N_{centres}$) for fixed dimension ($k$) and number of data-points ($N$):} In the construction of the grid the centres are not used. The BT is therefore independent of $N_{centres}$. In the QT we see that the curves are approximately separated by an order of magnitude, showing that it scales linearly with $N_{centres}$. This plot also shows that the optimal time is reached when $N_{cells}\sim2^7$, independently of the number of centres. However, the minimum TT is shifted to lower number of cells for a lower number of centres because the BT dominates over the QT.
    
\end{itemize}

\begin{figure*}
\includegraphics[width=\textwidth]{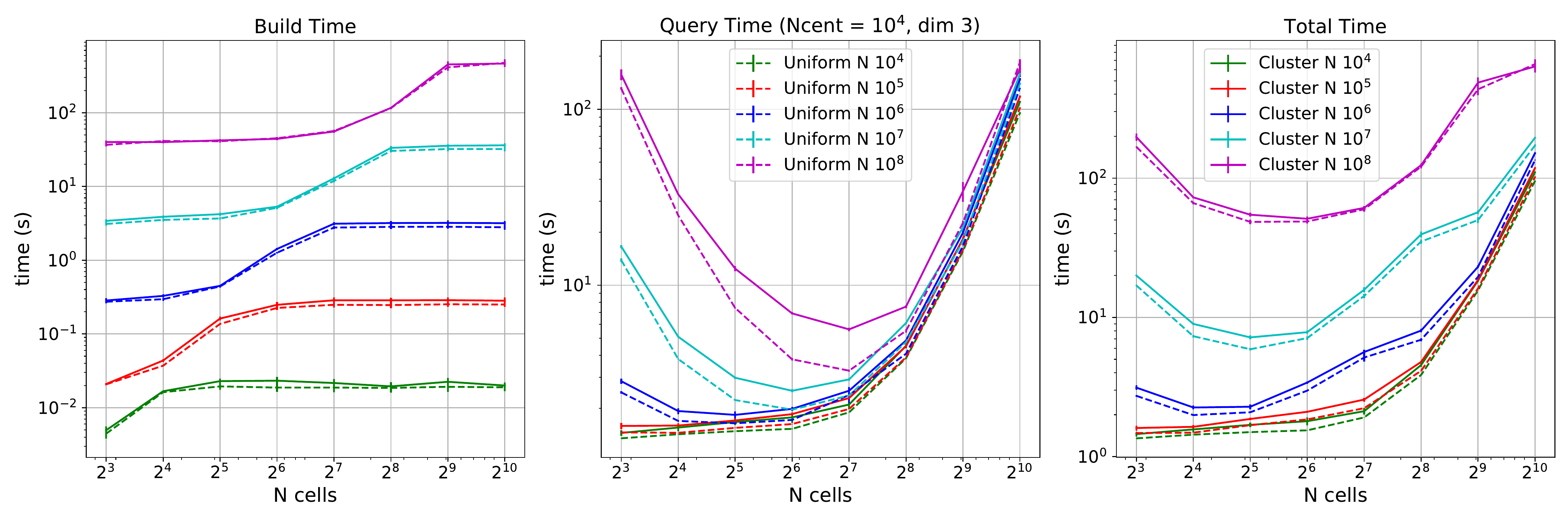}
\caption{Time benchmark of GriSPy in a gravitational N-body simulation using the \texttt{bubble\_neighbors} method with a search radii of 1\% of the box size. We compare the difference of considering uniformly distributed random centres against centres in a highly clustered region, i.e. dark matter halos.}
\label{fig:benchSimu}
\end{figure*}

We then study how GriSPy behaves in a more realistic scenario of a gravitational N-body simulation ($k=3$). We used the same simulation described in Section \ref{subsec:example}. For a direct comparison with the uniform case, particle positions are normalized with the box size to have values in the range (0, 1) in each dimension. As in the previous case, we used the \texttt{bubble\_neighbors()} method with a search radii of 0.01 and all distances are computed with the \textit{euclidean} metric. We restricted the analysis to the most relevant scenario, varying the number of data-points ($N$) for fixed number of centres ($N_{centres}$). Figure \ref{fig:benchSimu} shows the result of the scaling relation \textit{time vs. $N_{cells}$}. We also compare the difference of considering uniformly distributed random centres against centres in a highly clustered region, i.e. dark matter halos identified using a Friends of Friends algorithm with a standard linking length. The behaviour is exactly the same as the case (b) for the uniform set of points and there is no evident scaling factor influenced by the clustering. When considering different centres, however, there seems to be a slight increase in the QT for larger sets of points. Taking into account this analysis we consider that $N_{cells} = 2^6$ is an appropriate default value.

A similar benchmark analysis was carried out for the \texttt{shell\_neighbors()} and \texttt{nearest\_neighbors()} methods and their respective figures are included in \ref{appendix:a}. It should be noticed that the behaviour of the \texttt{shell\_neighbors()} method is identical to that shown by the \texttt{bubble\_neighbors()} method. The reason for this is that both algorithms are basically the same; the extra conditions evaluated in the \texttt{shell\_neighbors()} method to remove the inner region do not have an impact in the overall behaviour. In the case of the \texttt{nearest\_neighbors()} method, the behaviour is quite different due to the problem itself. Searching for a fixed number of neighbors is fundamentally a different problem than searching within a fixed radius. It should be noticed that solving the \textit{nth}-nearest neighbors by iteratively searching with \texttt{shell\_neighbors()} is not the best approach. In future releases of GriSPy, the \texttt{nearest\_neighbors()} method will be revised.

We run these tests in a node with the following specifications:

\begin{description}
    \item[CPU:] Intel Xeon CPU E5-2660v4 @ 2.00GHz 
    \item[RAM:] 128 GB DDR4 (1200-2001 MHz)
    \item[OS:] CentOS Linux 7 (Core) 64bits
    \item[Software:] \textit{Python} 3.7.5, \textit{NumPy} 1.17.3 and \textit{SciPy} 1.3.1
\end{description}

\subsection{Short comparison with similar projects} \label{subsec:cmp}

For the near neighbors search, GriSPy is based on the partition and indexing of the space through a regular grid. However, as we mentioned earlier, there are other solutions (in addition to brute force) for this purpose. For example, the best known packages make use of binary trees to address the search. In particular, the Scipy library \citep{Virtanen2020} implements the KDTree algorithm in Cython, cKDTree\footnote{\url{https://docs.scipy.org/doc/scipy/reference/generated/scipy.spatial.KDTree.html\#scipy.spatial.KDTree}}; while Scikit-learn \citep{scikit-learn,sklearn_api} also incorporates a BallTree\footnote{\url{https://scikit-learn.org/stable/modules/generated/sklearn.neighbors.BallTree.html\#sklearn.neighbors.BallTree}} scheme.

By contrasting the two comparable methods (search of the k nearest neighbor and all points within a given radius), the three classes exhibit a very similar ease of use, being able to deal with N-dimensional data.

Specifically, GriSPy presents almost the same utilities exposed by BallTree, except for the possibility of individually selecting the number of neighbors to search around each particular centre. However, GriSPy incorporates other additional features, such as protecting the original construction instance from any corruption or using periodic conditions of up to $1$ boxsize, individually adjustable on each axis.

On the other hand, cKDTree presents almost all these characteristics, extending the periodicity up to $n$ boxsizes, and incorporates some more as the possibility of using multiprocesses in the search. However, unlike the previous two classes which implement several metrics of Euclidean and non-Euclidean geometries, cKDTree is limited to use only a Minkowski p-norm, where p can vary in each search, and the user can not provide a custom distance function. Other additional search schemes are implemented by cKDTree, but among them is not the shell query method that distinguishes GriSPy. Full details of these comparisons with the most popular alternative packages can be found in \ref{appendix:b}.

We present in Figure \ref{fig:benchCodes} the time comparison of GriSPy against cKDTree and BallTree using the N-body simulation detailed in the previous section. For simplicity we consider only dark matter halos as centres and use the \texttt{bubble\_neighbors} method with a search radius of 1\% of the box size. In order to achieve a fair comparison we choose in every case the default configuration of each package. This means 64 $N_{cells}$ for GriSPy and no periodicity settings for any package. We also run the queries in a single process. The time benchmark analysis of each package shows that BallTree and cKDTree behave in an extremely similar way, probably due to similar schemes in their tree implementation. When considering the BT of each package, we can see that they are independent of the number of centres, as expected. However, the BT of GriSPy starts to slow when reaching approximately the same number of data-points as grid cells ($N_{cells}^3$), and becomes faster than the other packages at about $10^6$ data-points. The QT plot shows that GriSPy is slower than the other packages as would be expected. However, we notice that the QT of GriSPy grows at a slower rate compared to cKDTree and BallTree, reaching a difference of less than an order of magnitude when dealing with large data sets. The total time invested in building and querying shows that cKDTree and BallTree are faster than GriSPy for data sets of up to $\sim 10^7$, beyond that number GriSPy performs better. Furthermore, the slope of the curves indicates that GriSPy grows slower than the other methods and for data sets larger than $10^8$ the difference would still favour GriSPy.

From this analysis we think that the scenario in which GriSPy is a suitable solution to the NNS problem is when a query on a large data set ($\gtrsim 10^7$) is needed and also when the build step needs to be computed many times. Finally, some important points should be noticed. First, cKDTree methods return only the indices of neighbors for fixed-radius queries. If the user also needs the distance further computations are required. Second, BallTree has no periodic boundary conditions implemented. However, we decided to use it in the analysis for a complete comparison.

In the end, since these three packages are free and implemented in Python, all of them can take advantage of the Python scientific-stack synergy. It should be noticed that there are other astronomy related packages that perform neighbor searches, such as \textit{halotools} \citep{Hearin2017}, but we do not focus on them since the NNS problem is not their main purpose.

% ---------- COMPARISON benchmark
\begin{figure*}
\includegraphics[width=\textwidth]{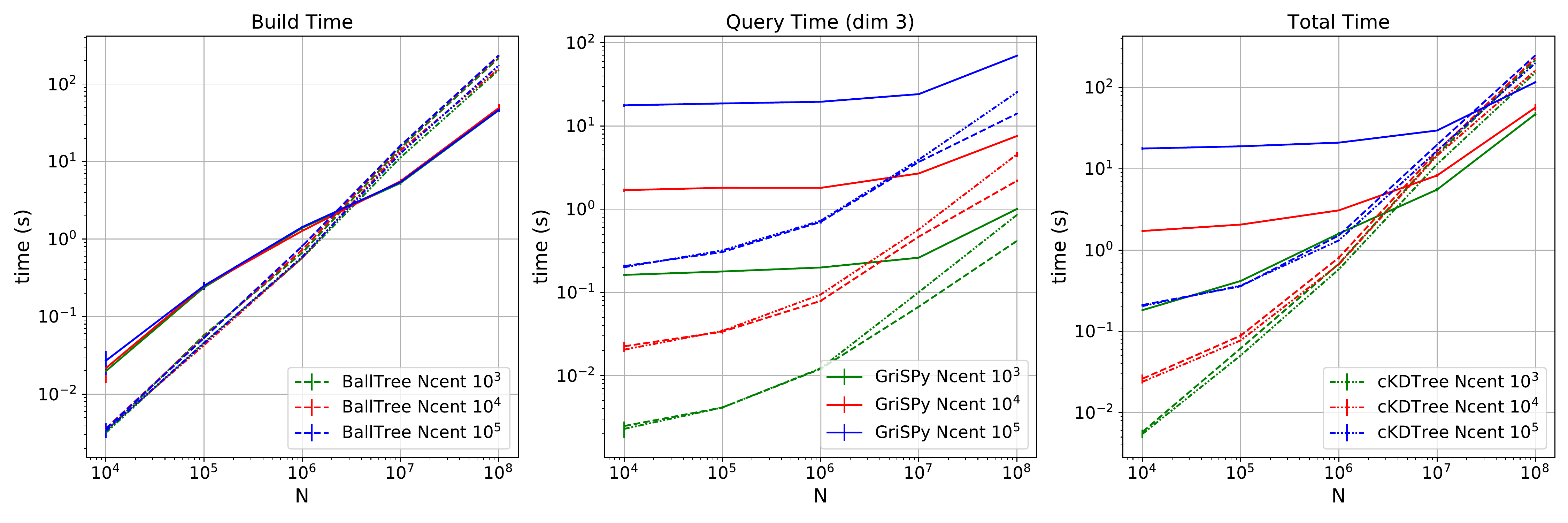}
\caption{Time comparison of GriSPy, cKDTree and BallTree in a gravitational N-body simulation using the \texttt{bubble\_neighbors} method with a search radii of 1\% of the box size. Each column represents the time spent in each step, from left to right: Build, Query and Total time. For clarity the legend specifying the package and number of centres is split across the three subplots. In all cases the error bars are the standard deviation of 10 realizations.}
\label{fig:benchCodes}
\end{figure*}

% =============================================================================
% Conclusions
% =============================================================================

\section{Conclusions}
In this paper we presented the first version of GriSPy: Grid Search in Python, a module for fast nearest neighbors searches. This algorithm indexes a set of \textit{k}-dimensional points in a regular grid or hypercube. Through a simple math operation every point is assigned an integer lattice that points to its corresponding cell. Then a hash table is constructed to save this information for later queries. In this first installment we provide the following types of query: \texttt{bubble\_neighbors()}, to find neighbors within a given radius; \texttt{shell\_neighbors()}, to find neighbors within given lower and upper radius; and \texttt{nearest\_neighbors()}, to find the \textit{n}-th nearest neighbors. We also implemented the following features: possibility of working with periodic boundary conditions in each dimension; individual query radius can be provided for each centre; three distance functions of interest in astronomy can be used (\textit{euclidean, haversine} and \textit{Vincenty}), and the possibility of providing a custom distance function.

\subsection{Caveats and future work}
The reader may have noticed that this first version of GriSPy has some limitations. The most notable is the fact that both, build and query, are performed in a single process. Paralellization is currently being developed, however to deliver the most efficient implementation further work is required.

Our prototypes make use of the \textit{Joblib} library \citep{varoquaux2009joblib}, which provides a unified interface to access thread-based and processes-based parallelism. On the other hand and if necessary, extensions to deploy \textit{Joblib} processes on distributed computing platforms such as \textit{Dask} and \textit{Spark} are available \citep{rocklin2015dask, spark2018apache}. It is important to note that parallelism is not the only  option for improvements since there are technologies such as the just-in-time compiler \textit{Numba} \citep{lam2015numba}, and the possibility of writing a critical code to some lower-level language like \textit{Cython} is always available \citep{behnel2011cython}.

Another aspect to improve is the algorithm behind the \texttt{nearest\_neighbors()} method, which has proven to be suboptimal. The main reason behind this is the fact that the entire cell-technique scheme was thought to have a high performance in fixed-radius queries and not in $n$-th nearest neighbors searches. Nevertheless, new ideas will be tested to deliver a practical method.

Finally, future releases of GriSPy will include new implementations such as new distance metrics, methods to return only counters instead of distances and indices, the possibility of computing two-point and three-point correlation functions, conditional $n$-th nearest neighbor queries (i.e. find the $n$-th nearest neighbors within a subset of data points that satisfy a given condition, for example a difference in magnitude: $|m_{points} - m_{centre}| < 3$).

% =============================================================================
% References
% =============================================================================

\bibliographystyle{aa}
\bibliography{grispy}

% =====================================================================================
% APPENDIX A
% =====================================================================================

\appendix
\onecolumn

\section{Complementary time-benchmark analysis}
\label{appendix:a}
In this section we include the figures corresponding to the benchmark analysis of the \texttt{shell\_neighbors} and \texttt{nearest\_neighbors} methods.
See \autoref{subsec:bench} for a full discussion.

% ---------- SHELL benchmark
\begin{figure*}[!h]
\includegraphics[width=\textwidth]{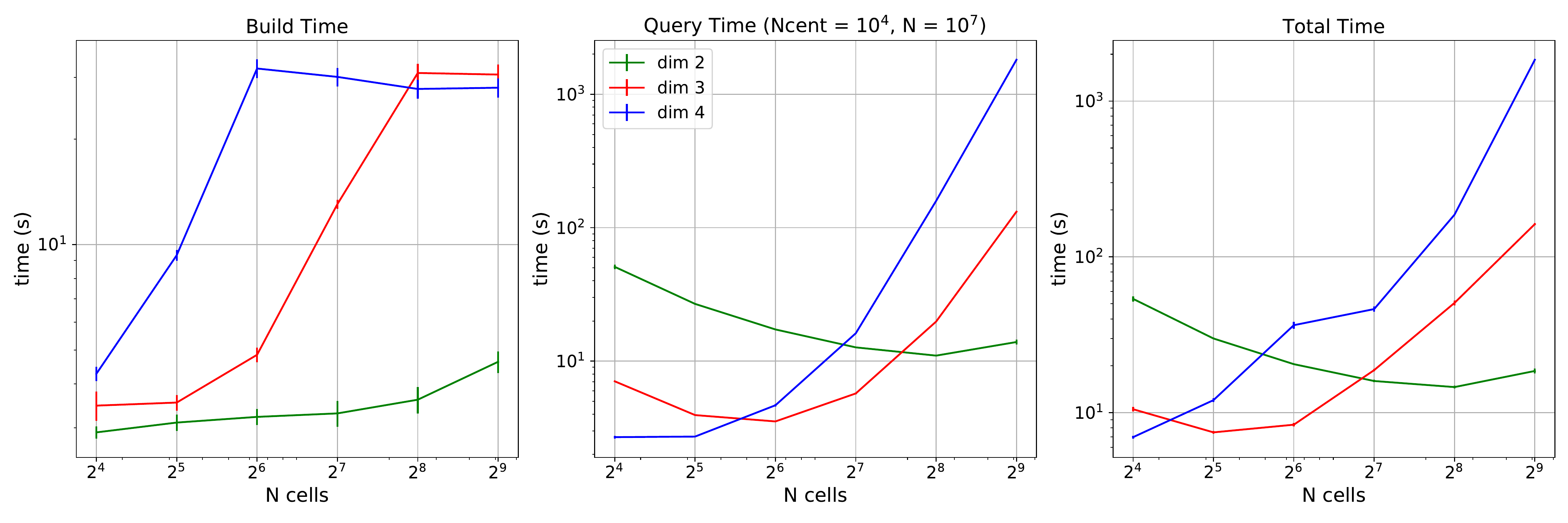}
\includegraphics[width=\textwidth]{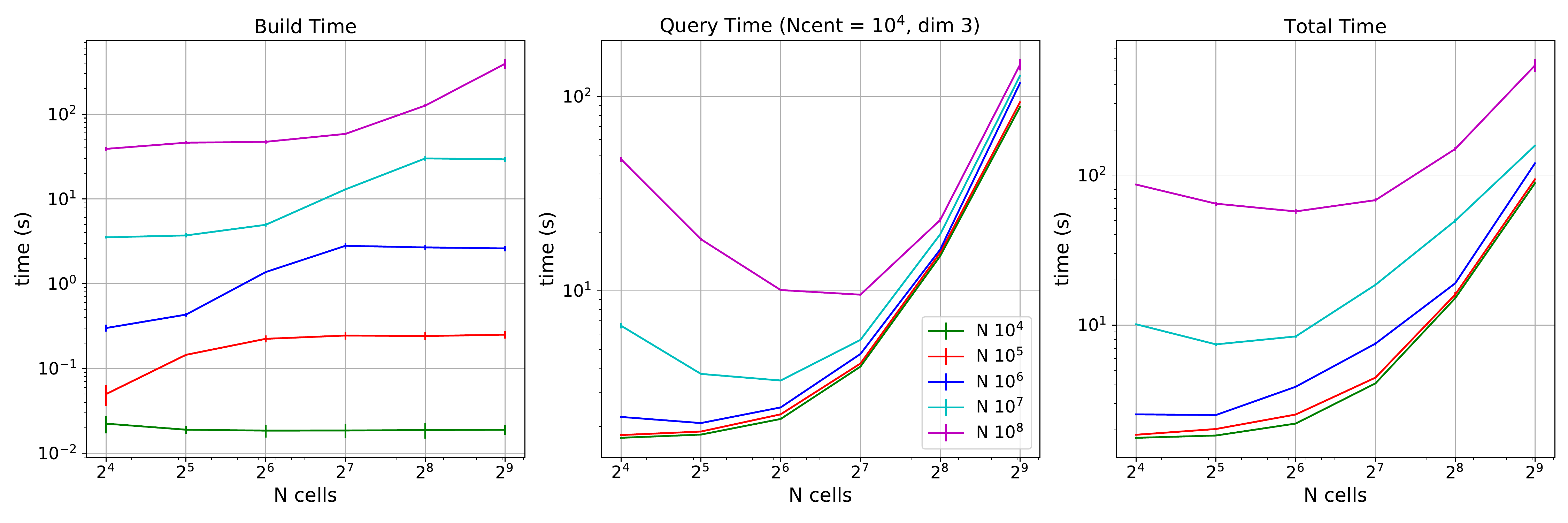}
\includegraphics[width=\textwidth]{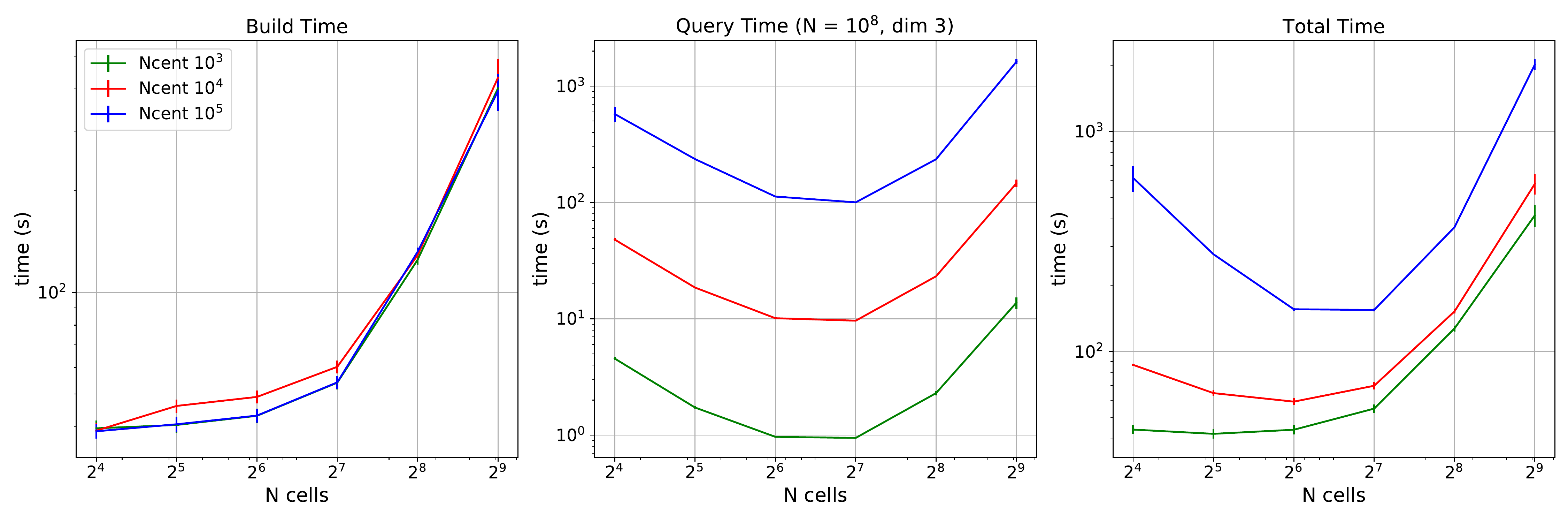}
\caption{Time benchmark of GriSPy in a uniform distribution of points using the \texttt{shell\_neighbors} method with lower and upper search radii of 1\% and 2\% of the box size, respectively. Each column represents the time spent in each step, from left to right: Build, Query and Total time. First row: varying dimension for fixed number of data-points ($N$) and centres ($N_{centres}$). Second row: varying number of data-points ($N$) for fixed dimension and number of centres ($N_{centres}$). Third row: varying number of centres ($N_{centres}$) for fixed dimension and number of data-points ($N$). In all cases the error bars are the standard deviation of 10 realizations.}
\label{fig:benchSHELL}
\end{figure*}

% ---------- NEAREST benchmark
\begin{figure*}
\includegraphics[width=\textwidth]{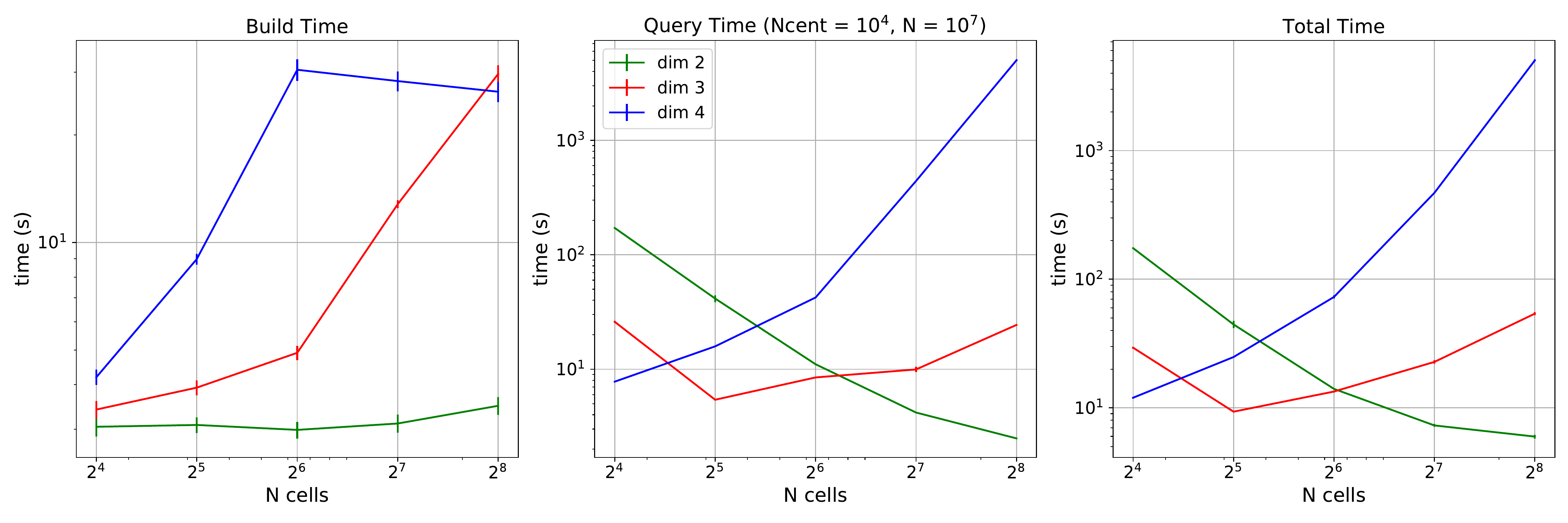}
\includegraphics[width=\textwidth]{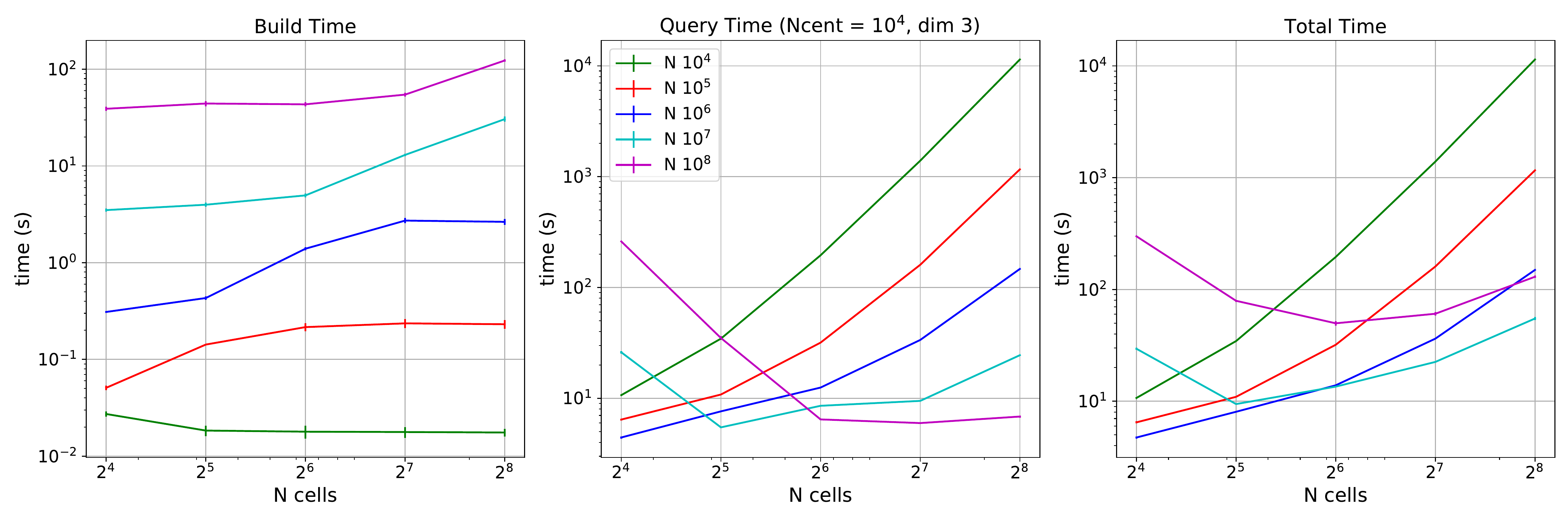}
\includegraphics[width=\textwidth]{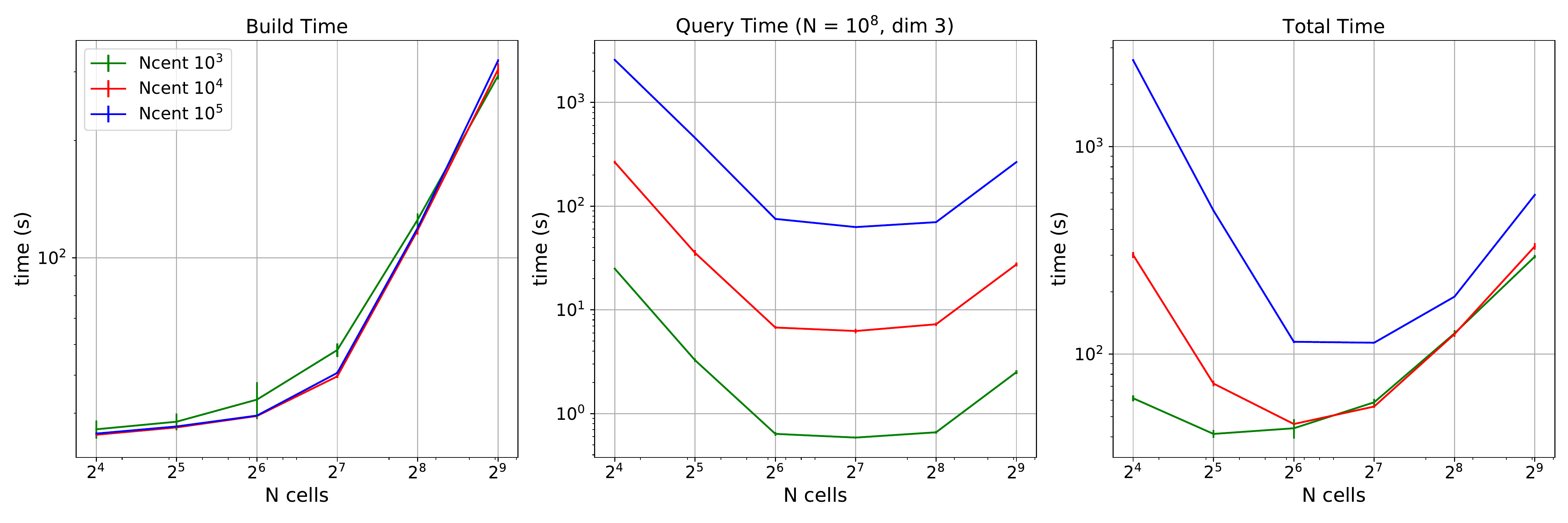}
\caption{Time benchmark of GriSPy in a uniform distribution of points using the \texttt{nearest\_neighbors} method. The number of neighbors is set to 128 in every case. Each column represents the time spent in each step, from left to right: Build, Query and Total time. First row: varying dimension for fixed number of data-points ($N$) and centres ($N_{centres}$). Second row: varying number of data-points ($N$) for fixed dimension and number of centres ($N_{centres}$). Third row: varying number of centres ($N_{centres}$) for fixed dimension and number of data-points ($N$). In all cases the error bars are the standard deviation of 10 realizations.}
\label{fig:benchNEAREST}
\end{figure*}

% =====================================================================================
% APPENDIX B
% =====================================================================================

\newpage
%\onecolumn

% \newcommand{\specialcell}[2][c]{%
%   \begin{tabular}[#1]{@{}c@{}}#2\end{tabular}}

\section{Comparative table for Nearest-Neighbors search packages}
\label{appendix:b}

% In the following table we summarize some of the most recognized Nearest-Neighbors implementations.
% %
% See \autoref{subsec:cmp} for a full discussion.

% \begin{table}[h]
% \centering
% \caption{NN Software Packages} \label{tab:sw_cmp}
% \resizebox{1.\textwidth}{!}{
% \begin{tabular}{l|llllll}
% \toprule
%      \textbf{Package} & \textbf{Language} &  \textbf{Algorithm} & \textbf{License} & \textbf{Homepage} & \textbf{Reference} \\
%      \midrule
     
%      GriSPy & 
%      Python & 
%      Grid search & 
%      \textsc{MIT} &	
%      \url{https://grispy.readthedocs.io} & 
%      \citep{} \\
     
%     \bottomrule
% \end{tabular}
% }
% \end{table}

In the following table we summarize some of the most recognized Near Neighbors search exact implementations in Python.
See \autoref{subsec:cmp} for a full discussion.

\begin{table}[h]
\centering
\caption{NN Software Packages} \label{tab:sw_cmp}
% \resizebox{1.\textwidth}{!}{
\begin{threeparttable}
\begin{tabular}{l|ccc}
\toprule
\textbf{Class}   &\textbf{cKDtree} & \textbf{BallTree}     & \textbf{GriSPy}\\
\midrule
\midrule
\textbf{Package} & \textbf{scipy}  & \textbf{scikit-learn} & \textbf{grispy}  \\
Module           & spatial         & neighbors             & -                \\
Version          & 1.3.3           & 0.22                  & 0.0.4            \\
\midrule 
Indexing structure & Binary tree   & Binary tree           & Fixed grid     \\
Data dimension     & N             & N                     & N              \\
Periodicity        & $n$ lbox      & No                    & $1$ lbox \vspace{0.2cm}\\

Distance metrics   & Minkowski p-norm & Internal or user defined   & Internal or user defined   \\
                   & Euclidean        & Euclidean \& Non-euclidean & Euclidean \& Non-euclidean \vspace{0.2cm}\\
                   
Copy data          & Yes           & No                    & Yes            \\
Multiprocessing    & Yes           & No                    & No             \\
\midrule
\multicolumn{4}{c}{Search methods} \\
\midrule
\textbf{bubble\tnote{+}} & \texttt{query\_ball\_point}  & \texttt{query\_radius}  & \texttt{bubble\_neighbors} \vspace{0.1cm}\\
search radii     & Multiples           & Multiples      & Multiples \\
return sorted    & Opt                 & Opt            & Opt  \\
count only       & Opt                 & Opt            & No   \\
return dist      & No                  & Opt            & Ever \vspace{0.2cm} \\
\textbf{k-NN\tnote{*}}   & \texttt{query}       & \texttt{query}          & \texttt{nearest\_neighbors} \vspace{0.1cm}\\
search neighbors & Multiples           & Multiples      & Single \\
return sorted    & Ever                & Opt            & Ever   \\
return dist      & Ever                & Opt            & Ever   \\
dist upper bound & Opt                 & No             & No   \vspace{0.2cm}\\
\textbf{Others\tnote{\#}} & \texttt{count\_neighbors}    & -              & \texttt{shell\_neighbors} \\
                 & \texttt{query\_ball\_tree}   & & \\
                 & \texttt{query\_pairs}        & & \\
\bottomrule
\end{tabular}
% }
\begin{tablenotes}\footnotesize
\item[+] Find all points within given distances of each centre.
\item[*] Find the k nearest-neighbors for each centre.
\item[\#] Query methods that are not comparable between classes. See packages documentation for more details.
\end{tablenotes}
\end{threeparttable}
\end{table}

\end{document}